\documentclass[twocolumn,showpacs,preprintnumbers,amsmath,amssymb]{revtex4-1}

\pdfoutput=1
\usepackage[pdftex]{graphicx}
\usepackage[caption=false]{subfig}

\usepackage{dcolumn}
\usepackage{bm}

\usepackage{color}
\usepackage{amstext}

\begin{document}

\title{Magnetic phase diagram of the Hubbard model in the Lieb lattice}

\author{J. D. Gouveia, R. G. Dias}
\affiliation{Departamento de F\'{\i}sica, I3N, Universidade de Aveiro, Campus de Santiago, Portugal}%

\date{\today}

\begin{abstract}
We study the mean-field phase diagram of the repulsive Hubbard model in the Lieb lattice. Far from half-filling, the most stable phases are paramagnetism for low on-site interaction $U/t$ and ferromagnetism for high $U/t$, as in the case of the mean-field phase diagram of the square lattice Hubbard model obtained by Dzierzawa [\onlinecite{Dzierzawa1992}].
At half-filling, the ground state was found to be ferrimagnetic [a $(\pi,\pi)$ spiral phase], in agreement with a theorem by Lieb [\onlinecite{Lieb1989}]. The total magnetization approaches Lieb's prediction as $U/t$ becomes large. As we move away from half-filling, this ferrimagnetic phase becomes a $(q_1,q_1)$ spiral phase with $q_1 \approx \pi$ and then undergoes a series of first-order phase transitions, $(q_1,q_1) \rightarrow (q_1,q_2) \rightarrow (q_1,0)$, with $q_2 \approx \pi/2$, before becoming ferromagnetic at large $U/t$ or paramagnetic at low $U/t$.
\end{abstract}


\maketitle

\section{Introduction}
\label{}

The Hubbard model is one of the most studied models in the area of strongly correlated electron systems \cite{Bednorz1986,RevModPhys.70.897}. However, it remains unsolved for dimensionality larger than one. For the one-dimensional (1D) case, the exact solution is given by the Bethe Ansatz \cite{Bethe}, while in the case of two dimensions (2D), the solution is known only in some limiting cases or by means of approximations, such as mean-field. The fermionic Hubbard model in a square lattice has long been known to display antiferromagnetism (AF) at half-filling \cite{Dias1992}. However, away from half-filling, the ground state magnetic ordering is still an open problem \cite{Marder2000}.

Extensions of the Hubbard model to 2D decorated lattices also show interesting features, such as flat band ferromagnetism (F) \cite{Lieb1989} and Dirac cones \cite{Wallace1947}. These decorated 2D lattices fall into three classes: Lieb's \cite{Lieb1989}, Mielke's \cite{Mielke1992} and Tasaki's \cite{Tasaki1992}. The pursuit for metallic ferromagnetism has motivated the search of crystal structures matching these decorated lattices. However, there are experimental obstacles, such as the lifting of the flat-band degeneracy by the Jahn-Teller effect or the difficulty in controlling the filling of the lattice. An alternative experimental approach is to study quantum dot arrays with these geometries \cite{Tamura2000}. Decorated lattices can also be realized by manipulating cold atoms in optical lattices \cite{Goldman2011}.

\begin{figure}[t]
\centering
\includegraphics[height=4cm]{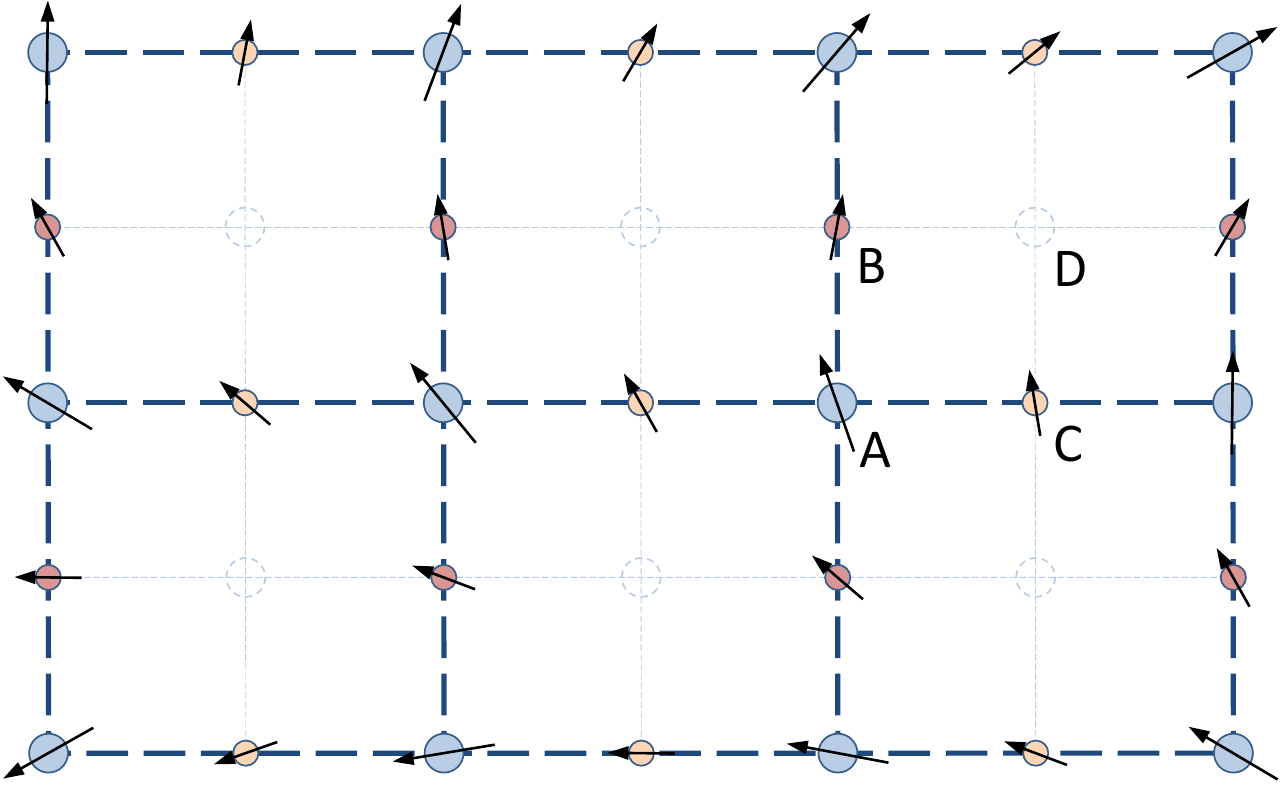}
\caption{A square lattice can be divided into four sublattices A, B, C and D. The circles represent atomic nuclei and the arrows represent spins. The Lieb lattice can be obtained by removing one of the sublattices.}
\label{fig-Lieb_lattice}
\end{figure}

Here, we study one example of a 2D decorated lattice, the Lieb lattice, i.e., a line-centered square lattice \cite{Wang2014}. This kind of lattice can be obtained from the usual 2D square lattice by removing a quarter of its atoms (see Fig. \ref{fig-Lieb_lattice}). Each unit cell contains one atom of each kind: A, B and C.
As a matter of fact, real materials can have their atoms arranged in a fashion resembling the Lieb lattice. Examples include the well-known high-$T_c$ superconductors with weakly coupled CuO$_2$ planes \cite{Emery1987,Scalettar1991}, such as La$_{2-x}$Sr$_x$CuO$_4$ and YBa$_2$Cu$_3$O$_7$, which can be studied using the perovskite lattice, a three-dimensional (3D) generalization of the Lieb lattice \cite{Weeks2010}.

Exact results for magnetism in the Lieb lattice are known. For example, an important theorem proven by Lieb \cite{Lieb1989} states that bipartite lattices (lattices with two sublattices, A and B, such that each site on sublattice A has its nearest neighbors on sublattice B, and vice versa) whose unit cell contains a different number of each kind of atom, have ferromagnetic ground states at half-filling. This is the case of the Lieb lattice \cite{Tamura2001}, as each unit cell contains one A atom and two B-like atoms. One common argument is that these states are in fact ferrimagnetic \cite{Mielke1993}, in the sense that although each sublattice is ferromagnetic, the full lattice is antiferromagnetic, but the magnetization is finite due to the different number of atoms in each sublattice. This contrasts with the antiferromagnetic ordering of the square lattice Hubbard model in this limit. Note that Lieb's theorem only mentions the total magnetization per unit cell, not on-site magnetization amplitudes, which can be calculated using numerical methods, such as mean-field. This has been done for the multi-layer Lieb optical lattice at half-filling \cite{Noda2014}.

In this work, we use a mean-field approach to compute the magnetic phase diagram of the Lieb lattice as a function of the average electron density $n$ and Hubbard interaction $U$, thus going away from both half-filling and the tight-binding limit. The allowed magnetic phases are paramagnetism and spin spiral phases \cite{S.Sarker1991}. Ferro- and ferrimagnetism can be obtained as particular cases of spiral phases. Note that we do not consider spatial phase separation. In order to find such regions in the phase diagram, one needs to use the chemical potential as an independent variable \cite{Langmann2007,P.A.Igoshev2010,Schumacher1983}, rather than using the particle density.

The tight-binding Hamiltonian of the Lieb lattice, $H_t$, is given by \cite{Nita2013}
\begin{equation}
\begin{array}{l}
t \sum\limits_{x=1}^{L_x} \sum\limits_{y=1}^{L_y} \left[ (A_{x,y}^\dag B_{x,y}   + A_{x,y}^\dag C_{x,y}   + H.c.) \right. \\
\phantom{aaa} \left. + (A_{x,y}^\dag B_{x,y-1} + A_{x,y}^\dag C_{x-1,y} + H.c.) \right] .
\end{array}
\label{eq-Hamiltonian_Lieb_tb_complete}
\end{equation}
$L_x$ ($L_y$) is the number of unit cells along the $x$ ($y$) direction. The hopping terms in the first line are intra-unit cell and the remaining ones are inter-unit cell. The eigenvalues of $H_t$ originate three energy bands, one of which is flat. The dispersion relation for periodic boundary conditions is
\begin{equation}
\varepsilon_{\pm} = \pm 2 t \sqrt{\cos^2 \frac{k_x}{2} + \cos^2 \frac{k_y}{2}} ,
\label{eq-Lieb_dispersionrelation}
\end{equation}
for the non-flat energy bands, where $k_{\alpha} = 2 \pi n_{\alpha} /L_{\alpha}$ with $n_{\alpha} = 0,1,\cdots ,L_{\alpha}$ and $\alpha \in \{ x,y \}$. The flat band is $L_x L_y$-fold degenerate with zero energy. The one-particle localized states associated with the flat bands can be written as
\begin{equation}
\left| \text{loc} ; x,y \right\rangle = \frac{1}{2} \left( B_{x,y}^\dagger - C_{x,y}^\dagger + B_{x,y-1}^\dagger - C_{x-1,y}^\dagger \right) \left| \text{vac} \right\rangle .
\end{equation}
These states form a non-orthogonal basis of the flat band subspace.

\begin{figure}[t!]
\centering
\subfloat[ ]{\label{fig-Lieb_tb_dispersionrelation}\includegraphics[width=.18 \textwidth]{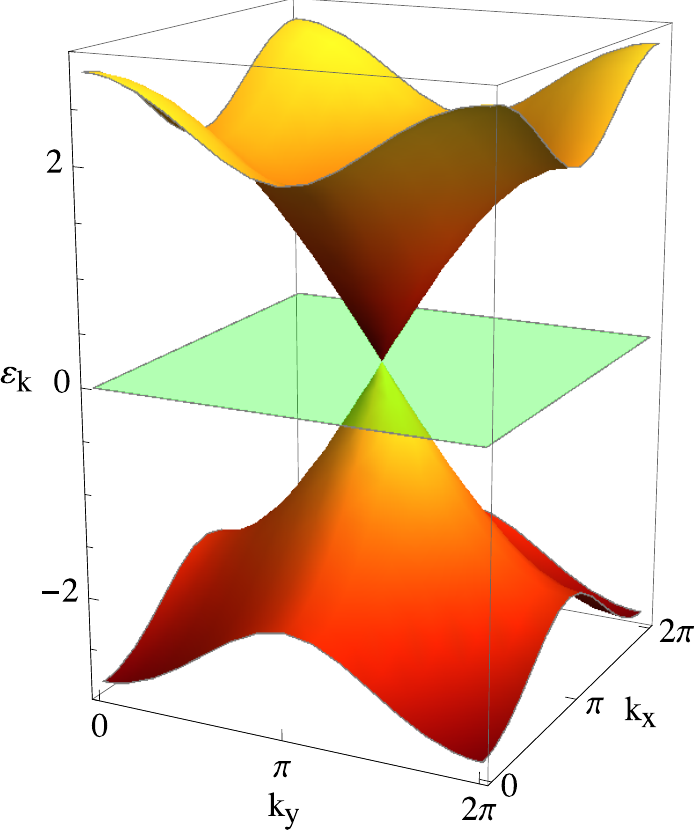}}
\subfloat[ ]{\label{fig-Lieb_tb_sublatticeocupation}\includegraphics[width=.32 \textwidth]{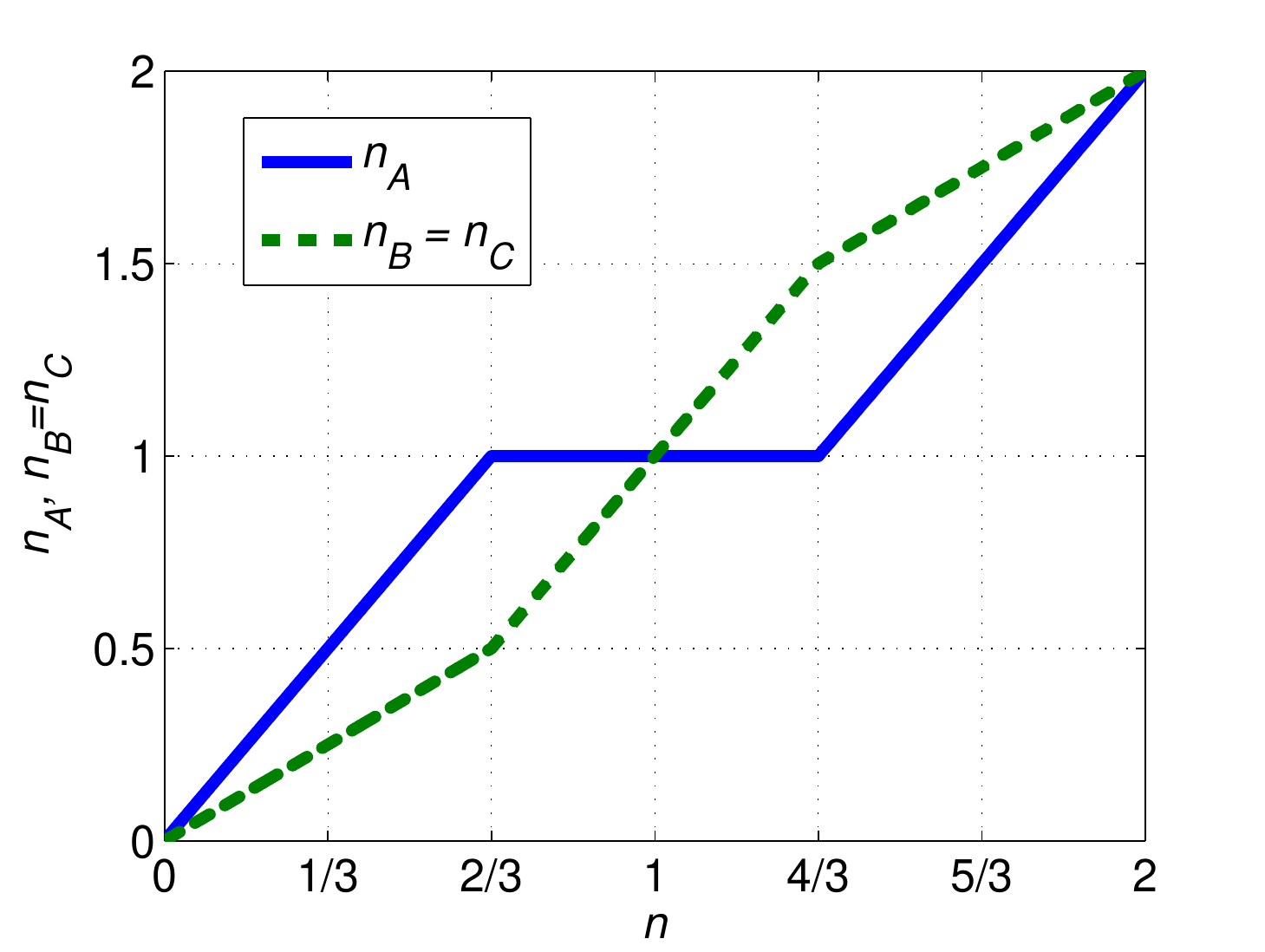}}
\caption{(a) Plot of the tight-binding dispersion relation of the Hubbard model in the Lieb lattice and (b) the respective particle density of each sublattice, A, B or C, as a function of the total particle density.}
\label{fig-Lieb_tb_E_and_n}
\end{figure}
The three tight-binding energy bands of the Lieb lattice energy bands are shown in Fig. \ref{fig-Lieb_tb_dispersionrelation}. At the point $(k_x,k_y) = (\pi,\pi)$, the three branches touch each other. Expanding the dispersion relation in Eq. \ref{eq-Lieb_dispersionrelation} around this point, we find the Dirac cones $\varepsilon^2 = t^2 (k_x^2 + k_y^2)$. The flat band is built up from B- and C-type orbitals in equal shares, while the lower and upper bands involve all three sublattices A, B and C.

The particle density of a sublattice equals the number of electrons at that sublattice divided by the number of atoms the sublattice comprises, or the number of unit cells,
\begin{equation}
\begin{array}{l}
n_A = \frac{N_A}{L_x L_y} \\
n_B = \frac{N_B}{L_x L_y} \\
n_C = \frac{N_C}{L_x L_y} .
\end{array}
\end{equation}
In the non-interacting limit and at zero temperature, the particle density on sublattices A, B and C as a function of the global particle density (number of electrons in the whole lattice divided by the total number of sites) is as plotted in Fig. \ref{fig-Lieb_tb_sublatticeocupation}. The plot can be interpreted as follows. Half of the probability density of the states in the lower dispersive band correspond to the sublattice A, while the other half is evenly distributed among sublattices B and C. Therefore, starting at $n = 0$, as we insert electrons in the system, half ``choose'' sublattice A, while the other half go to sublattices B or C. At $n = 2/3$, we reach the flat band at $\varepsilon=0$. At this point, all sites A are singly occupied, while sites B and C are quarter-filled. Any newly-added electrons will only go to sublattices B or C, because the flat band only comprises these two kinds of atoms and going to sublattice A would imply going to the upper dispersive band, which would lead to higher total energy. At $n = 4/3$, the flat band is completely filled, so that for $n > 4/3$ electrons occupy the upper dispersive band going to sites A or B/C at a ratio of 2:1, as in the lower dispersive band, up to the maximum filling $n_A = n_B = n_C = 2$.

In this work, we address magnetism in the Lieb lattice by considering a finite on-site Coulomb repulsion $U$ using a mean-field approach, and build a $n-U$ phase diagram. In the case of a square lattice, one assumes that the occupation number is the same in the whole lattice. Here, in the case of the Lieb lattice, we assume that the occupation number on each sublattice is the same as in the tight-binding limit, for any $U$ (see Fig. \ref{fig-Lieb_tb_sublatticeocupation}). This is the correct assumption for small $U/t$. Moreover, for large $U/t$, the results of Fig. \ref{fig-results_square_Lieb} remain qualitatively the same for $n_A = n_B = n_C = n$.

\section{Calculations}
\begin{figure}[tbph]
\centering
\includegraphics[height=4cm]{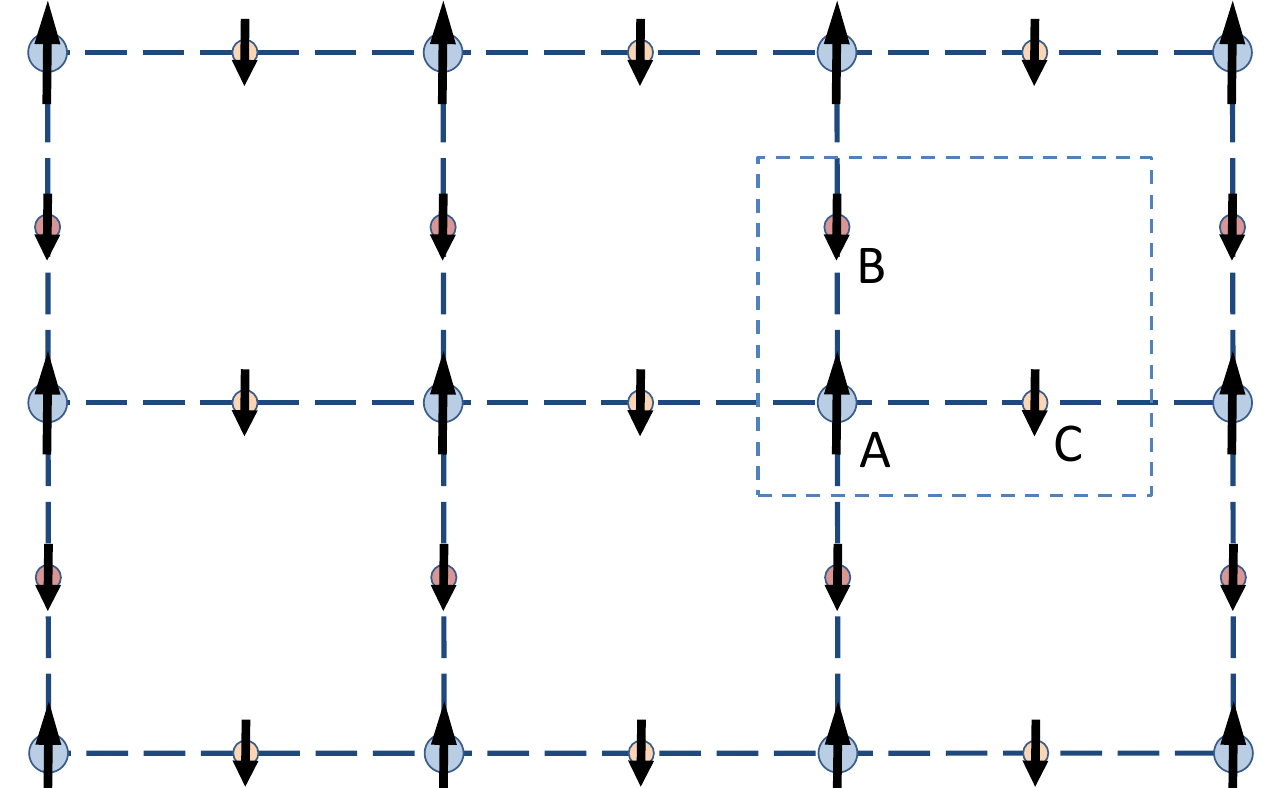}
\caption{According to our definition of $\vec{q}$ (the change in spin orientation between two consecutive lattice sites), a Lieb lattice with $\vec{q} = (\pi,\pi)$ is ferromagnetic within each sublattice. Moreover, because the magnetization amplitude $m$ is the same in every site and each unit cell has two spins in the same direction and one spin in the opposite direction, the total spin per unit cell is nonzero, as predicted by Lieb \cite{Lieb1989}.}
\label{fig-Fsublattice_AFglobal}
\end{figure}

The interaction term of the Hubbard Hamiltonian is
\begin{equation}
H_U = U \sum\limits_{\text{sites}} n_\uparrow n_\downarrow,
\label{eq-Hamiltonian_interaction}
\end{equation}
that is, the on-site Coulomb repulsion $U$ times the number of double occupancies in the lattice. Applying the mean-field approximation to the Hubbard Hamiltonian gives single-particle energies given by the eigenvalues of the $6 \times 6$ single-particle Hamiltonian $H_{\text{MF}}$ \cite{Dzierzawa1992,A.Singh1992,Gouveia2014},
\begin{widetext}
\begin{equation}
\left( \begin{array}{cccccc}
\frac{Un_A}{2}   & -t(1+e^{i k_y})         & -t(1+e^{i k_x})         & -\frac{mU}{2}           & 0                         & 0                         \\
-t(1+e^{-i k_y}) & \frac{Un_B}{2}          & 0                       & 0                       & -\frac{mU}{2} e^{-i q_y}  & 0                         \\
-t(1+e^{-i k_x}) & 0                       & \frac{Un_C}{2}          & 0                       & 0                         & -\frac{mU}{2} e^{-i q_x}  \\
-\frac{mU}{2}    & 0                       & 0                       & \frac{Un_A}{2}          & -t(1+e^{i (k_y+2q_y)})    & -t(1+e^{i (k_x+2q_x)})    \\
0                & -\frac{mU}{2} e^{i q_y} & 0                       & -t(1+e^{-i (k_y+2q_y)}) & \frac{Un_B}{2}            & 0 \\
0                & 0                       & -\frac{mU}{2} e^{i q_x} & -t(1+e^{-i (k_x+2q_x)}) & 0                         & \frac{Un_C}{2}
\end{array} \right) ,
\label{eq-HMF}
\end{equation}
\end{widetext}
plus the diagonal term
\begin{equation}
\frac{U L_{\text{uc}}}{4} (3m^2-n_A^2-n_B^2-n_C^2) .
\label{eq-diagonal_term}
\end{equation}
This is a generalization of the Hamiltonian obtained in previous studies of the 2D square lattice Hubbard model \cite{Dzierzawa1992,A.Singh1992,Gouveia2014}, which did not allow for different occupations in the sublattices. The magnetic phase of the system is defined by two order parameters: the vector $\vec{q}$ and the number $m$, as in the works by Dzierzawa \cite{Dzierzawa1992} and Singh \cite{A.Singh1992}. The vector $\vec{q} = (q_x,q_y)$ defines the orientation of the spins. For example, $q_x = 0$ is a ferromagnetic phase along the $x$ direction, $q_y = \pi$ represents antiferromagnetism along the $y$ direction, and other values of $q_x$ or $q_y$ give spin spiral phases. The paramagnetic phase is $\vec{q}$-degenerate and is characterized by zero magnetization amplitude. The magnetization amplitude $m$ can be identified as the amplitude of the spin spiral wave,
\begin{equation}
\langle \vec{S}_{\vec{r}} \rangle = \frac{m}{2} \left( \cos(\vec{q} \cdot \vec{r}) , \sin(\vec{q} \cdot \vec{r}) , 0 \right) ,
\end{equation}
and appears during the mean-field calculations, when computing averages such as
\begin{equation}
\langle A_\uparrow^\dagger A_\downarrow \rangle = \langle S_A^+ \rangle = \langle S_{A,x} + i S_{A,y} \rangle = \frac{m}{2} e^{i \vec{q} \cdot \vec{r}_A} ,
\end{equation}
for sublattice A. Fig. \ref{fig-Fsublattice_AFglobal} shows what the configuration of the Lieb lattice looks like when $\vec{q} = (\pi,\pi)$ and $m$ is finite.

From this point forward, we consider $t=1$, so that $U$ is given in units of $t$. It is important to remark that, experimentally, although we cannot directly control the value of $U$ (the on-site interaction), we can control the ratio $U/t$, for example by applying pressure on the sample.

\section{Results and discussion}

The $n-U$ phase diagram is computed in the following way. For each point $(n,U)$, the number of electrons $N$ is well defined, so that we can add the lowest $N$ mean-field energies and find the total energy of the system. By numerically minimizing this total energy (using, for instance, the algorithm in Ref. \cite{Lagarias1998}) with respect to $q_x$, $q_y$ and $m$, we find the values of these three magnetic order parameters which lead to the ground state for this pair $(n,U)$. Repeating this process for all desired pairs, one obtains the phase diagram.

\subsection{Magnetization for high $U/t$}
\begin{figure}[t]
\centering
\includegraphics[width=.4\textwidth]{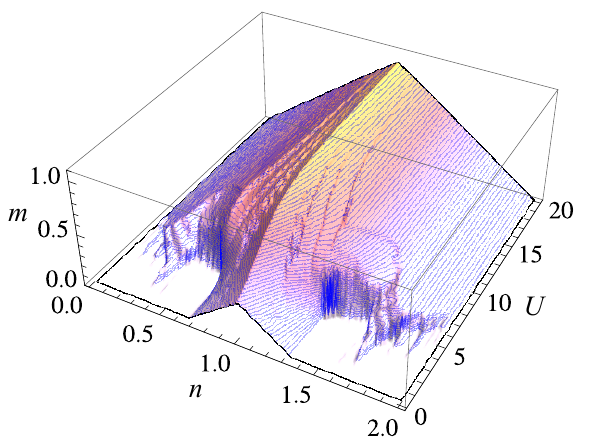}
\caption{Mean-field ground state magnetization ($m$) of the Lieb lattice as a function of $n$ and $U$. The plot is very similar to that of the 2D square lattice. The most noticeable difference between the two is that, while the square lattice has zero $m$ in the vicinity of the point $(n,U) = (1,0)$, the Lieb lattice has finite $m$ in this region of the diagram, more specifically between $n=2/3$ and $n=4/3$.}
\label{fig-plotm}
\end{figure}

The plot in Fig. \ref{fig-plotm} shows the mean-field ground state magnetization amplitude $m$ as a function of $n$ and $U$. This result is similar to that of the square lattice in most regions of the diagram. Indeed, for high $U$, the magnetization is proportional to $n$ between $n=0$ and $n=1$, and proportional to $2-n$ between $n=1$ and $n=2$, reflecting particle-hole symmetry.

This proportionality can be justified analytically in the following way. For very high $U$, the tight-binding terms of the mean-field Hamiltonian given by Eq. \ref{eq-HMF} are but a small perturbation, which can be neglected as a first approximation. In this case, all sublattices become equivalent, implying $n_A=n_B=n_C=n$, with $t=0$. Diagonalizing the new Hamiltonian gives two flat bands. A three-fold degenerate band at $\frac{U}{2} (n-m)$ and a three-fold degenerate band at $\frac{U}{2} (n+m)$. Distributing the electrons in the bands and adding up their energies, one obtains the total energy of the system by adding the diagonal term $\frac{3UL_{\text{uc}}}{4} (m^2-n^2)$ (see Eq. \ref{eq-diagonal_term}), so that having positive $m$ yields the same bands as negative $m$. Let us assume $m>0$. For $n \in [0,1]$, electrons occupy only the lowest (degenerate) energy band, with energy $\frac{U}{2} (n-m)$. The total energy is then given by
\begin{equation}
\frac{3UL_{\text{uc}}}{4} (m^2-n^2) + \frac{U}{2} \sum\limits_N (n-m) .
\end{equation}
The result of the summation is simply $N(n-m) = 3nL_{\text{uc}}(n-m)$. Minimizing this with respect to $m$ gives the expected result $m=n$. Performing an analogous calculation assuming $n>1$ yields the relation $m = 2-n$, i.e., the other half of the plot in Fig. \ref{fig-plotm} for high $U$.

\subsection{Magnetization for low $U/t$}

The results for the magnetization amplitude $m$ in the limit $U \rightarrow 0$ can be explained using first-order perturbation theory. Let us denote by $H_0$ (the unperturbed Hamiltonian) the tight-binding terms of Eq. \ref{eq-HMF}, that is, Hamiltonian $H_{MF}$ with $U=0$. Its eigenvalues are
\begin{equation}
\begin{array}{c}
\pm 2 t \sqrt{\cos^2 \frac{k_x}{2} + \cos^2 \frac{k_y}{2}} , \\
\pm 2 t \sqrt{\cos^2 \left( \frac{k_x}{2} + q_x \right) + \cos^2 \left( \frac{k_y}{2} + q_y \right) } ,
\end{array}
\label{eq-eigenvalues_tb_Lieb}
\end{equation}
and two coincident flat bands at $\varepsilon = 0$. Using the interaction terms of Hamiltonian \ref{eq-HMF} as a perturbation yields, to first order, two key results. Firstly, the flat bands are split into two non-degenerate nearly flat bands. One of them is shifted to positive energy by an amount proportional to $mU$, while the other is shifted to negative energy, by the same amount, at each point of the Brillouin zone. Secondly, the four non-flat bands are shifted by $\frac{U}{4} (n_A + n_B)$. These two conclusions allow us to predict the behaviour of $m$ near $U=0$. For the following calculations, one must keep in mind that the diagonal term in Eq. \ref{eq-diagonal_term} is also to be accounted for.

We begin by filling up the lower bands (which correspond to the bands with minus signs in Eq. \ref{eq-eigenvalues_tb_Lieb}), distributing the particles among the sublattices according to Fig. \ref{fig-Lieb_tb_sublatticeocupation}. For $n$ lower than $2/3$, it is best to keep $m=0$ because, up to first order, the energy of the two lower-energy bands (associated with the Hamiltonian $H_0$) does not depend on $m$, and having finite $m$ would only increase the total energy due to the diagonal term in Eq. \ref{eq-diagonal_term}. As the total particle density reaches $n \approx 2/3$ (getting closer to $2/3$ as $U$ approaches $0$), we start to fill the nearly flat bands at $\varepsilon = 0$. This is the point at which a finite $m$ can be used to lower the energy of one of the flat bands, thus lowering the total energy of the system. After the lower flat band has been filled (note that finite $U$ induces some modulation of the flat bands, but the argument is valid for small perturbations), we are at $n=1$ and start filling the upper flat band. Now, it becomes advantageous to lower the value of $m$, so as to reduce the energy of this band. Finally, at $n \approx 4/3$, only the two higher-energy dispersive bands remain empty and between $n=4/3$ and $n=2$, the value of $m$ goes back to zero, for the same reason as when filling the two lowest-energy bands.

Let us now compare these assertions with our numerical results in Fig. \ref{fig-plotm}. At small $U$ and far from half-filling (outside the $n$ interval $[2/3 ; 4/3]$), the ground state of the system is paramagnetic ($m=0$), coinciding with the square lattice result (see Fig. \ref{fig-diagram_square}). On the other hand, inside the interval $n \in [2/3 ; 4/3]$, the square lattice becomes paramagnetic (except at exactly $n=1$, where it is antiferromagnetic, and in a very small region around $n=1$, where a spiral phase arises; the width of this region goes to zero as $U/t \rightarrow \infty$) while our result suggests that the Lieb lattice has a magnetic ordering other than paramagnetism. To know which ordering it is, one needs to look at the results for $q_x$ and $q_y$.

\subsection{Magnetic ordering}

\begin{figure*}[tbp]
\centering
\subfloat[ ]{\label{fig-diagram_square}\includegraphics[width=.7 \textwidth]{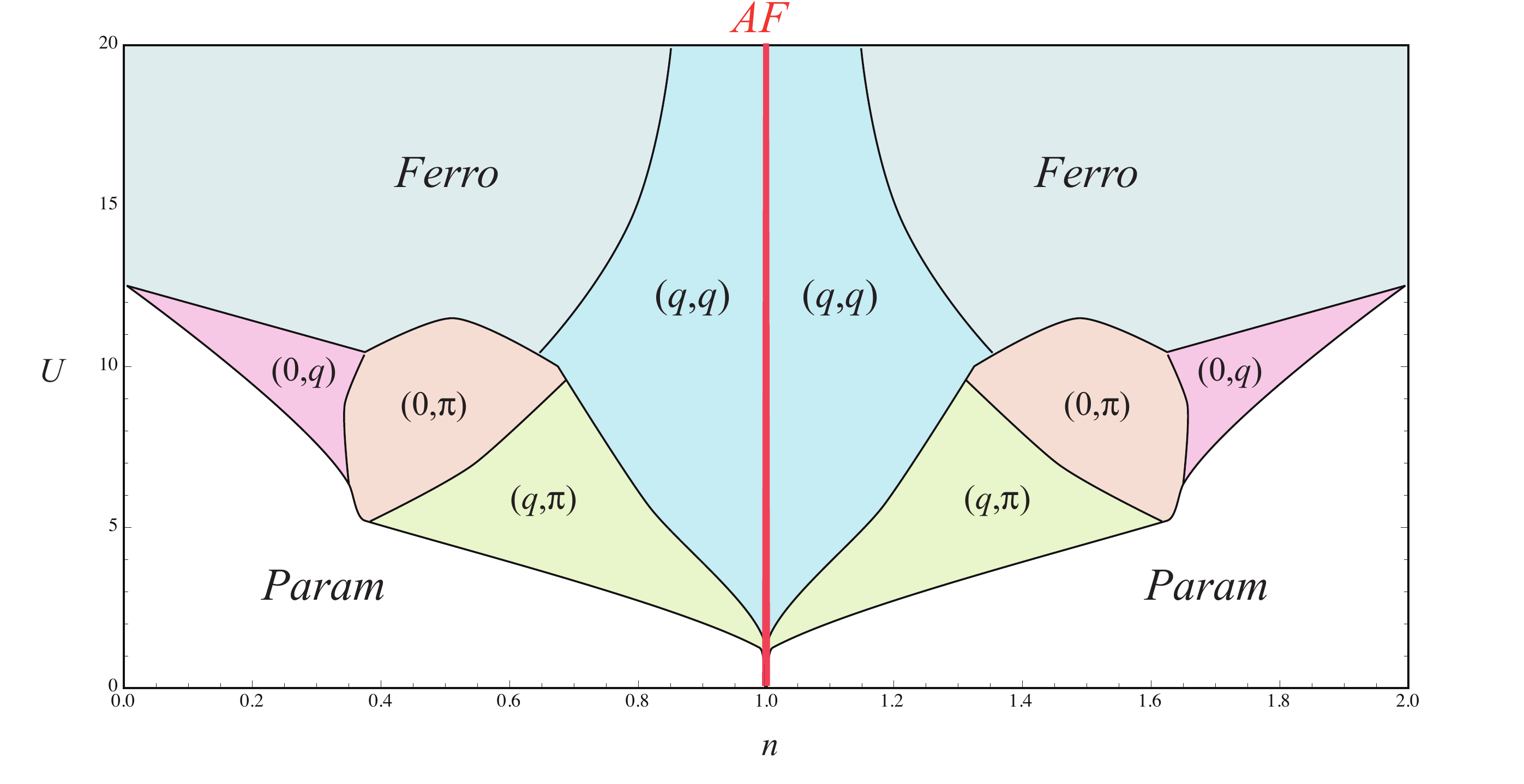}}\\
\subfloat[ ]{\label{fig-diagram_Lieb}\includegraphics[width=.7 \textwidth]{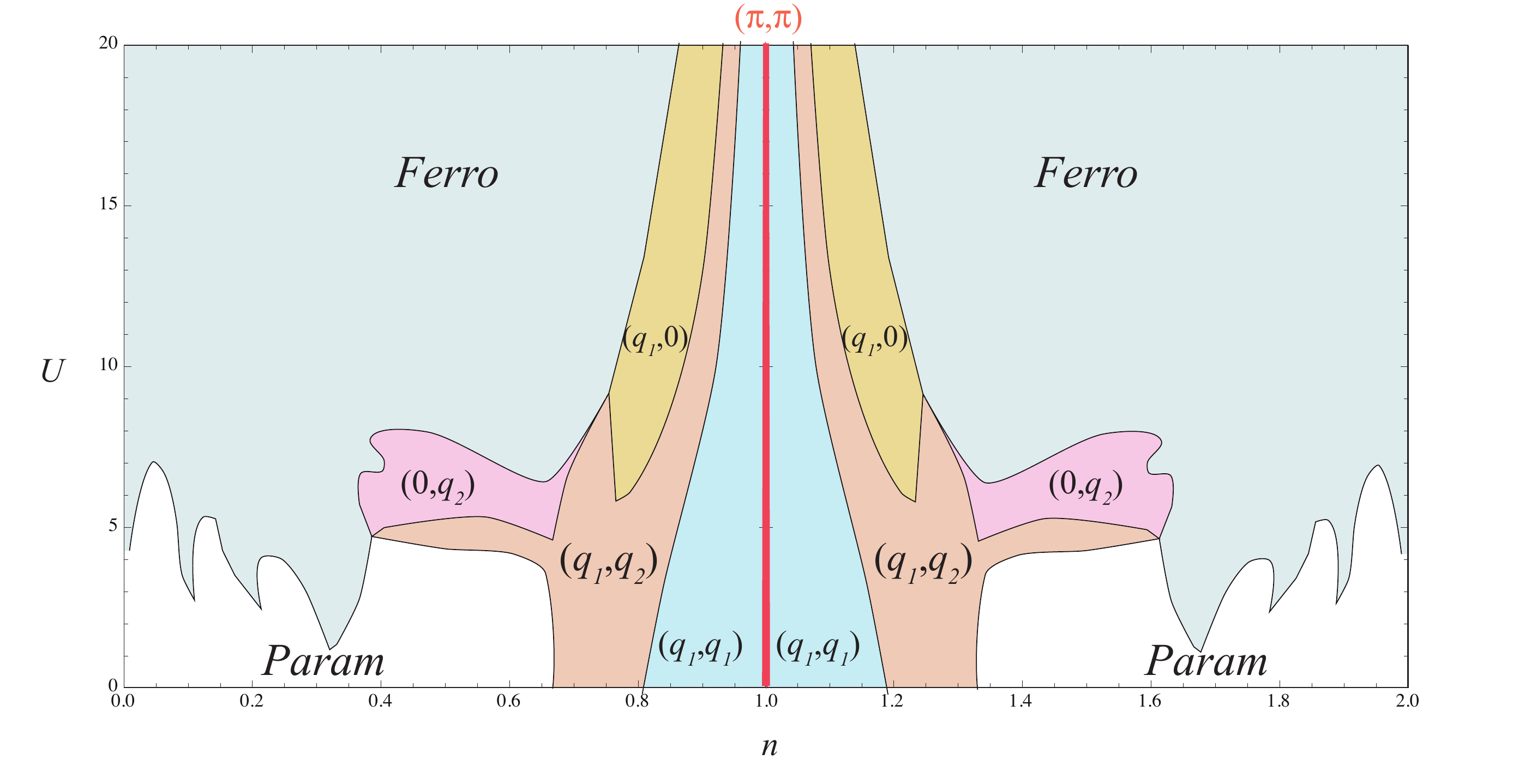}}
\caption{Mean-field magnetic phase diagrams of (a) the square lattice and (b) the Lieb lattice. In the case of the square lattice, the value of $q$ varies continuously in the range $[0,\pi]$, in each region labelled as such. In the case of the Lieb lattice, $q_1 \approx \pi$ and $q_2 \approx \pi/2$, and the transitions between regions labelled using $q_1$ or $q_2$ are discontinuous.}
\label{fig-results_square_Lieb}
\end{figure*}

Fig. \ref{fig-results_square_Lieb} shows both the mean-field magnetic phase diagram of the Lieb lattice (bottom plot) and that of the square lattice (top plot), for comparison. The phase diagrams were computed using the independent variables $n$ and $U$, in the range $(n,U) \in [0,2] \times [0,20]$, and were obtained by joining the results for the three order parameters: $q_x$, $q_y$ and $m$. Near $n=0$ and $n=2$, the system is ferromagnetic for large $U$ and paramagnetic for low $U$, like the square lattice, albeit with a wider ferromagnetic region. At intermediate $U$ and $n$ near 0.5 or 1.5, the system displays a $(0,q_2)$ spiral phase, characterized by $q_2 \approx \pi/2$.

The spiral phase characterized by $\vec{q} = (\pi,\pi)$ only occurs at exactly $n=1$, for any $U$, as in the square lattice. In the latter, this would be called antiferromagnetism. Nonetheless, in the Lieb lattice, a $(\pi,\pi)$ phase should be identified with ferrimagnetic ordering \cite{Lieb1989,Mielke1993} (see Fig. \ref{fig-Fsublattice_AFglobal}). Indeed, the spin-spin correlation in a $(\pi,\pi)$ phase is ferromagnetic in each sublattice, but antiferromagnetic between different sublattices. The total spin per unit cell is finite, because $m$ is finite at half-filling (see Fig. \ref{fig-plotm}) and the number of sites per unit cell is odd.

When slightly doped away from half-filling ($0.95 \lesssim n \lesssim 1.05$), both $q_x$ and $q_y$ continuously deviate from $\pi$ and become a $(q_1,q_1)$ phase with $q_1 \approx \pi$. This area becomes narrower in the $n$ direction as $U$ grows larger. This phase can be interpreted as a $(\pi-\delta,\pi-\delta)$ phase with small $\delta$, that is, a local (looking at only a few unit cells) ferrimagnet with a slow modulation in the direction of spins along the lattice. At large $U$, when further doped, the system undergoes a first-order phase transition from $\vec{q} \approx (\pi,\pi)$ to $\vec{q} \approx (q_1,q_2)$ with $q_2 \approx \pi/2$, reflecting local antiferromagnetic correlations in the $x$ direction, and sublattice-wise antiferromagnetism in the $y$ direction. In other words, each sublattice is ferromagnetic in the $x$ direction and antiferromagnetic in the $y$ direction. If doped even further away from half-filling, two more first-order phase transitions occur: first to $(q_1,0)$ and finally to $(0,0)$ (ferromagnetism). At regular intervals in $n$ (namely 0.11, 0.22, 0.33 and their symmetric counterparts), we find ferromagnetic dips into the paramagnetic region. These can most likely be explained using the symmetry of the lattice and higher-order corrections.

\section{Conclusion}

In summary, we have computed and analysed the $n$-$U$ mean-field magnetic phase diagram of the Lieb lattice, and compared it to that of the square lattice. Far from half-filling, the two phase diagrams display ferromagnetism [$\vec{q} = (0,0)$] for high $U$ and paramagnetism ($m = 0$) for low $U$, while at exactly half-filling (one electron per lattice site) the ground state is a $(\pi,\pi)$ spiral phase for both lattices.

Although the diagrams coincide at $n=1$, it is close to that line that their most remarkable differences arise. In fact, at large $U$, as we move away from half-filling [(the $(\pi,\pi)$ phase)], the Lieb lattice undergoes three first-order phase transitions $(\pi,\pi) \rightarrow ~(\pi,\pi/2) \rightarrow ~(\pi,0) \rightarrow (0,0)$, unlike in the case of the square lattice, where the transition from antiferromagnetism to ferromagnetism is continuous in $\vec{q}$. On the other hand, near the tight-binding limit and within the interval $n \in [2/3,4/3]$, the magnetization of the Lieb lattice is finite and the ground state is characterised by spin spirals, contrasting with the paramagnetic ordering of the square lattice in this region of the diagram.

Our numerical results are in agreement with a theorem by Lieb \cite{Lieb1989}, which applies to Hubbard models which comprise sublattices with different number of sites (in our case, we have twice as many B/C sites as we have A sites). The theorem states that the ground state of such a system at half-filling is ferrimagnetic \cite{Mielke1993}. According to our results, the ground state at half-filling is characterized by $\vec{q} = (\pi,\pi)$ and finite $m$, which translates into ferromagnetic sublattices and finite total spin on each unit cell (see Fig. \ref{fig-Fsublattice_AFglobal}), which is qualitatively consistent with the theorem. According to this theorem, however, the total magnetization per unit cell in the Lieb lattice at half-filling should be 1 for any $U$. Our mean-field approach yields that value as $U$ grows large but deviates from 1 at low $U$ (see Fig. \ref{fig-plotm}). On the other hand, this theorem is also applicable to a square lattice Hubbard model if one divides the lattice into two sublattices. This has been done before \cite{Gouveia2014} with a square lattice divided into two sublattices, A and B, with the same number of sites each. In consonance with the aforementioned theorem by Lieb, this square lattice has zero total spin per unit cell at half-filling, for any $U$, regardless of $m$ being zero or not. Therefore, it stands to reason to conjecture that the mean-field calculations performed for the square lattice also return wrong values for $m$ at low $U$, even though the correct values cannot be deduced from Lieb's theorem, as it only predicts the total spin per unit cell.

The disparity between our mean-field results at half-filling and the prediction of Lieb's theorem may be due to two important restrictions that we imposed in order to simplify our calculations. Firstly, we assumed that the occupation numbers for any $U$ remain the same as in the tight-binding limit ($U=0$), and secondly, we assumed that the magnetization is the same on every sublattice. If it turns out that these two assumptions are indeed the reason for the discrepancy, that is, if the Lieb's theorem can be satisfied in a mean-field approach applied to this paper's model, albeit with more relaxed constraints, such a result should be taken into account in any other mean-field study of interacting fermions in bipartite lattices or even more complex lattices whose unit cells contain more than two types of atoms. This is an open question that we intend to address in the future.

\section*{Acknowledgements}

R. G. Dias acknowledges the financial support from the Portuguese Science and Technology Foundation (FCT) through the program PEst-C/CTM/LA0025/2013.

J. D. Gouveia acknowledges the financial support from the Portuguese Science and Technology Foundation (FCT) through the grant SFRH/BD/73057/2010.

\bibliography{helix1}
\end{document}